\documentclass[8.5pt,twoside,twocolumn]{article}
\usepackage[super,sort&compress,comma]{natbib} 
\usepackage{mhchem}
\usepackage{times,mathptmx}
\usepackage{sectsty}
\usepackage{balance} 

\usepackage{graphicx} %eps figures can be used instead
\usepackage{lastpage}
\usepackage{cleveref}
\usepackage[format=plain,justification=raggedright,singlelinecheck=false,font=small,labelfont=bf,labelsep=space]{caption} 
\usepackage{fancyhdr}
\begin{document}

\thispagestyle{plain}
\fancypagestyle{plain}{
\fancyhead[L]{\includegraphics[height=8pt]{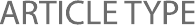}}
\fancyhead[C]{\hspace{-1cm}\includegraphics[height=20pt]{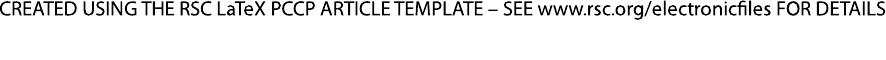}}
\fancyhead[R]{\includegraphics[height=10pt]{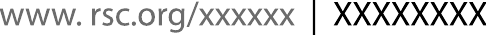}\vspace{-0.2cm}}
\renewcommand{\headrulewidth}{1pt}}
\renewcommand{\thefootnote}{\fnsymbol{footnote}}
\renewcommand\footnoterule{\vspace*{1pt}% 
\hrule width 3.4in height 0.4pt \vspace*{5pt}} 
\setcounter{secnumdepth}{5}

\makeatletter 
\def\subsubsection{\@startsection{subsubsection}{3}{10pt}{-1.25ex plus -1ex minus -.1ex}{0ex plus 0ex}{\normalsize\bf}} 
\def\paragraph{\@startsection{paragraph}{4}{10pt}{-1.25ex plus -1ex minus -.1ex}{0ex plus 0ex}{\normalsize\textit}} 
\renewcommand\@biblabel[1]{#1}            
\renewcommand\@makefntext[1]% 
{\noindent\makebox[0pt][r]{\@thefnmark\,}#1}
\makeatother 
\renewcommand{\figurename}{\small{Fig.}~}
\sectionfont{\large}
\subsectionfont{\normalsize} 

\fancyfoot{}
\fancyfoot[LO,RE]{\vspace{-7pt}\includegraphics[height=9pt]{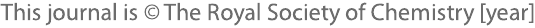}}
\fancyfoot[CO]{\vspace{-7.2pt}\hspace{12.2cm}\includegraphics{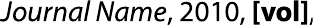}}
\fancyfoot[CE]{\vspace{-7.5pt}\hspace{-13.5cm}\includegraphics{RF}}
\fancyfoot[RO]{\footnotesize{\sffamily{1--\pageref{LastPage} ~\textbar  \hspace{2pt}\thepage}}}
\fancyfoot[LE]{\footnotesize{\sffamily{\thepage~\textbar\hspace{3.45cm} 1--\pageref{LastPage}}}}
\fancyhead{}
\renewcommand{\headrulewidth}{1pt} 
\renewcommand{\footrulewidth}{1pt}
\setlength{\arrayrulewidth}{1pt}
\setlength{\columnsep}{6.5mm}
\setlength\bibsep{1pt}

\twocolumn[
  \begin{@twocolumnfalse}
\noindent\LARGE{\textbf{Evaporative Deposition in Receding Drops$^\dag$}}
\vspace{0.6cm}

\noindent\large{\textbf{Julian Freed-Brown}}\vspace{0.5cm}
%Please note that \ast indicates the corresponding author(s) but no footnote text is required. 

\noindent\textit{\small{\textbf{Received Xth XXXXXXXXXX 20XX, Accepted Xth XXXXXXXXX 20XX\newline
First published on the web Xth XXXXXXXXXX 200X}}}

\noindent \textbf{\small{DOI: 10.1039/b000000x}}
\vspace{0.6cm}
%Please do not change this text.

\noindent \normalsize{We present a framework for calculating the surface density profile of a stain deposited by a drop with a receding contact line. Unlike a pinned drop, a receding drop pushes fluid towards its interior, continuously deposits mass across its substrate as it evaporates, and does not produce the usual ``coffee ring.'' For a thin, circular drop with a constant evaporation rate, we find the surface density of the stain goes as $\eta(r) \propto \left(\left(r/a_0\right)^{-1/2}-r/a_0\right)$, where $r$ is the radius from the drop center and $a_0$ is the initial outer radius. Under these conditions, the deposited stain has a mountain-like morphology. Our framework can easily be extended to investigate new stain morphologies left by drying drops.}
\vspace{0.5cm}
 \end{@twocolumnfalse}
  ]

%%%%%%%%%%%%%%%%%%%%%%%%%%%%%%
%%%%%%%%%%%%%%%%%%%%%%%%%%%%%%

%\begin{abstract} %REVTEX
%We present a framework for calculating the surface density profile of a stain deposited by a drop with a receding contact line. Unlike a pinned drop, a receding drop pushes fluid towards its interior, continuously deposits mass across its substrate as it evaporates, and does not produce the usual ``coffee ring.'' For a thin, circular drop with a constant evaporation rate, we find the surface density of the stain goes as $\eta(r) \propto \left(\left(r/a_0\right)^{-1/2}-r/a_0\right)$, where $r$ is the radius from the center of the drop and $a_0$ is the initial radius. Under these conditions, the deposited stain has a mountain-like morphology. Our framework can easily be extended to investigate new stain morphologies left by drying drops.
%\end{abstract}

% \maketitle %REVTEX

%%%%%%%%%%%%%%%%
% INTRODUCTION %
%%%%%%%%%%%%%%%%

%Footnotes
\footnotetext{\dag~Electronic Supplementary Information (ESI) available: [details of any supplementary information available should be included here]. See DOI: 10.1039/b000000x/}

%Please use \dag to cite the ESI in the main text of the article.
%If you article does not have ESI please remove the the \dag symbol from the title and the above footnotetext.

\footnotetext{\textit{Department of Physics and the James Franck Institute, University of Chicago, 929 E 57th Street, Chicago, IL 60637, USA. \\E-mail: jfreedbrown@uchicago.edu}}

%additional addresses can be cited as above using the lower-case letters, c, d, e... If all authors are from the same address, no letter is required

\section{Introduction}

Solute deposition from evaporating sessile drops is an important tool with varied applications. Evaporation-controlled deposition is used in colloidal self assembly,\cite{abkarian2004colloidal, nobile2009self, byun2010hierarchically, marin2011order, li2013macroscopic} electronics, \cite{kim2006direct, park2014flexible} particle segregation, \cite{monteux2011packing} and medical physics.\cite{brutin2011pattern} Currently studied effects on evaporative deposition include Marangoni flow,\cite{hu2006marangoni} substrate shape,\cite{xu2007evaporation, hong2009robust} and surface-bound colloids.\cite{yunker2011suppression} Here, we examine the impact of a receding contact line on solute deposition in an evaporating sessile drop.

Evaporation and changes in height force fluid flow within a drying drop and eventually govern the shape of its deposited stain. Often, the outer edge of the drop pins to the surface, giving rise to the ``coffee ring effect,'' \cite{deegan1997capillary, deegan2000contact} but there are many cases when the edge does not stay pinned throughout evaporation. Many studies examine the connection between stain morphology and the stick-slip behavior of the contact line. \cite{deegan2000pattern, yabu2005preparation, maheshwari2008coupling, frastia2011dynamical, mampallil2012control, zhang2014coffee} Furthermore, a recent study by Li \emph{et al}\cite{li2014solute} found that drops containing poly(ethylene glycol) recede during the majority of their evaporation and only show contact line pinning at early times. These drops form mountain-like deposits in the center of the drop instead of the usual ring morphology. Other experiments have also found unusual, mountain-like stain morphologies for receding drying drops.\cite{willmer2010growth, baldwin2012monolith} In these cases, the standard intuition from the coffee ring effect does not apply. When the drop's edge freely recedes, height changes near the edge of the drop are larger than height changes at the center, which pushes fluid radially inward as the drop evaporates. Furthermore, mass deposits when it reaches the receding edge of the drop, which means mass continuously deposits across the surface instead of the original outer edge.

Despite these differences, many of the ideas from the coffee ring effect can be extended to drops with receding contact lines. In this paper, we theoretically examine solute deposition in a thin, circular drop with a constant evaporation profile across its surface and a receding contact line. For this specific case, the surface density of the stain can be calculated analytically. To find the surface density of the deposited stain we
\begin{enumerate}
\item find the height profile as a function of time;
\item determine the fluid velocity from evaporation and changes in height;
\item solve for the trajectory of fluid parcels; and
\item track the evolution of masses bound by fluid parcels. 
\end{enumerate}
We find an exact formula for the surface density of the stain deposited from uniform evaporation. Our calculation serves as an illustrative example of the effects from a receding contact line and our method can easily be implemented numerically to handle more complex evaporation profiles.

%%%%%%%%%%%%%%%%%%%%%%
% THEORETICAL REGIME %
%%%%%%%%%%%%%%%%%%%%%%

\section{Theoretical Regime}

In our calculation, we consider small, circularly symmetric drops with slow dynamics. In these drops, surface tension dominates over gravitational, viscous, and inertial stresses. The drop height evolves quasi-statically and surface tension alone governs its shape. For drops with viscosity, density, and surface tension comparable to water, this regime corresponds to drop radii of a few millimeters and drying times of thousands of seconds. Note that these scales compare with scales found in experiments.\cite{deegan2000contact, li2014solute}

The drop surface meets the substrate with a contact angle $\theta_c$. Moving contact lines are generally characterized by an advancing and a receding contact angle controlled by substrate heterogeneity \cite{drelich1996effect} and evaporating drops are often observed to recede at fixed contact angles.\cite{li2014solute} Based on these observations, we examine the case where $\theta_c$ is constant.

We also work in the thin drop limit, where $\theta_c$ is small. In the thin drop limit, the mean curvature of the drop height is constant. Additionally, vertical velocities are negligible compared to radial velocities. For very thin drops, vertical length scales are much smaller than horizontal length scales and the diffusion length of solute over short times can be larger than typical vertical distances but smaller than typical horizontal distances. Then, we can assume that solute stays completely mixed in the vertical direction while horizontal diffusion is negligible. Furthermore, we adopt the limit where the solute is a dilute (its initial concentration, $\phi_0$, is small) and non-interacting. Under these assumptions, the solute acts as a passive tracer and is simply transported with the fluid's depth averaged velocity. We can neglect stratifying effects and any jamming or shock fronts that could be caused by solute interaction, which greatly simplifies the analysis. These assumptions were used by Deegan \emph{et al}\cite{deegan2000contact} to make useful predictions for pinned drops. We adopt the same conditions to facilitate comparison between the pinned and unpinned cases.

%%%%%%%%%%%%%%%%%%
% DROP EVOLUTION %
%%%%%%%%%%%%%%%%%%

\section{Drop Evolution}

The fluid velocity within the drop determines the final surface density of the deposited stain. The depth averaged velocity obeys a continuity equation because the mass of fluid is conserved.

%%%%%%%%%%%%%%
% CONTINUITY %
%%%%%%%%%%%%%%
\begin{figure}[t]
   \centering
   \includegraphics[width=.4\textwidth]{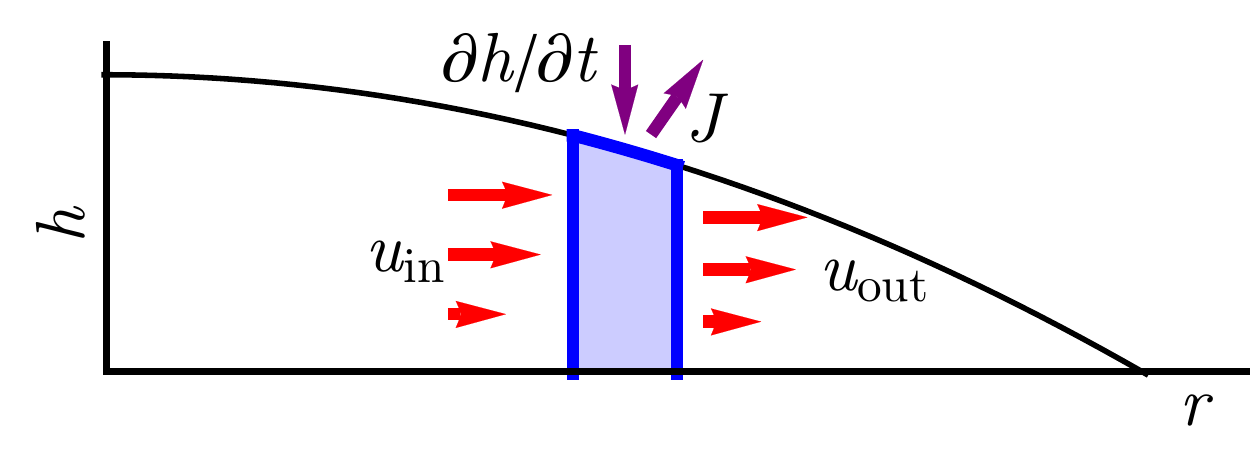}
   \caption{
   Radial vertical section of droplet indicated by black line representing height, $h$, vs radial distance, $r$. Surface evaporation $(J)$ and changes in height $(\partial h/\partial t)$ induce flow ($u$) within drop.
   }
   \label{fig:continuity}
\end{figure}
%%%%%%%%%%%%%%

Consider an annulus of fluid within the drop (See Fig.~\ref{fig:continuity}). As time evolves, the height profile decreases and acts as a local source of fluid flow $(\partial h/\partial t)$. Meanwhile, fluid  leaves the surface through evaporation, which acts as a local sink. The evaporation rate, $J(r,t)$, is the volume of fluid removed per unit surface area per unit time at radius $r$ and time $t$. Together, evaporation and the change in height act as a source in a continuity equation for the velocity field:
\begin{equation}
\nabla\cdot(h u) = -\frac{\partial h}{\partial t}- \left(1+\left(\frac{\partial h}{\partial r}\right)^2\right)^{1/2} J(r,t),
\end{equation}
where $hu$ is the fluid flux, $\partial{h}/\partial{t}$ is the change in height over time, and $(1+(\partial h/\partial r)^2)^{1/2}$ is the surface area element.

In the thin drop limit, $\frac{\partial h}{\partial r} \ll1$ and this equation simplifies to
\begin{equation}
\nabla\cdot(h u)=-\frac{\partial h}{\partial t} - J(r,t).
\label{eq:continuity}
\end{equation}
To find the fluid velocity, we first use surface tension to find the height profile at a given time. By balancing the total volume change of the height profile against the total evaporation rate, we also solve for the time dependence of the height profile. From there, we substitute $h$ into Eq.~2 to solve for the velocity.

% Finding the height profile
\subsection{Height Profile}
Surface tension determines the shape of the drop's surface. In the thin drop limit, the mean curvature of the surface $(\nabla^2 h)$ is constant and the drop is assumed to have a constant contact angle with the surface, $\theta_c$, as it evaporates. Together, these constraints determine the shape of the drop, but not its time dependence. 

\begin{figure}[t]
   \centering
   \includegraphics[width=.4\textwidth]{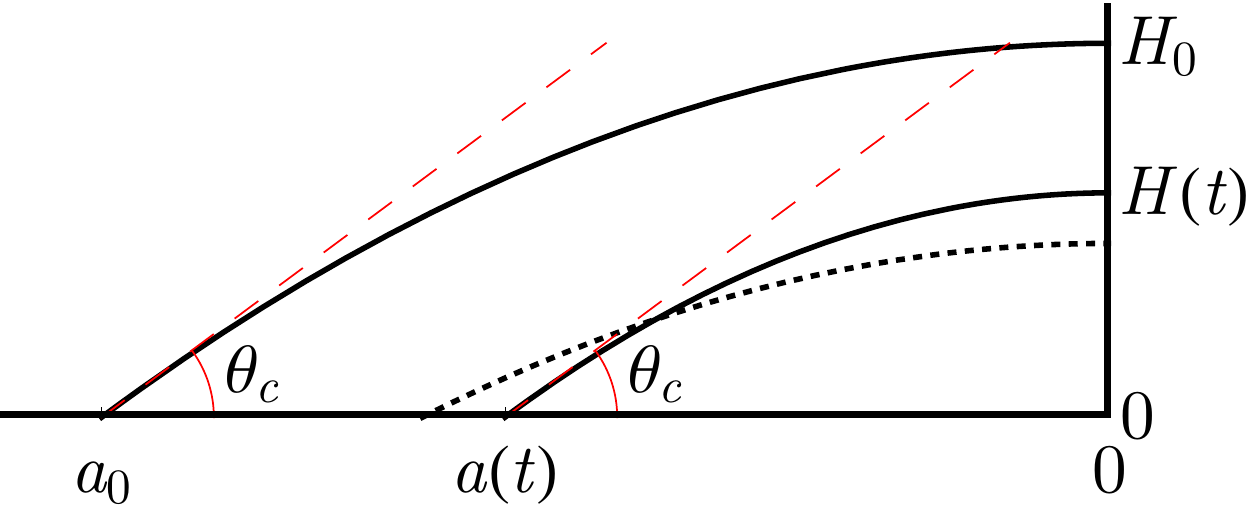}
   \caption{
		Height profile at two times, showing time evolution of height profile in evaporating drop. If height profile shifts vertically (dotted line), contact angle decreases. Fluid must retract as height decreases to maintain a constant contact angle.
   }
   \label{fig:height}
\end{figure}

In the thin drop limit, the height of an axisymmetric drop is given by
\begin{equation}
	h(r,t) = H(t) \left(1-\frac{r^2}{a^2(t)}\right),
	\label{eq:h}
\end{equation}
where the height at the center of the drop, $H(t)$, and the maximum radius of the drop, $a(t)$, are both functions of time. Furthermore, $H(t)$ is proportional to $a(t)$ because the drop's contact angle with the surface is constant---that is, $\left.\frac{\partial h}{\partial r}\right|_{r=a(t)}$ is constant. While $H(t)$ and $a(t)$ have some unknown time dependence, they vary in time together (See Fig.~\ref{fig:height}).

%%%%%%%%%%%%%%%%%%
% HEIGHT PROFILE %
%%%%%%%%%%%%%%%%%%
%%%%%%%%%%%%%%%%%%

Evaporation from the drop's surface determines the exact form of the time dependence. The total evaporation must exactly balance the drop's volume change since evaporation conserves the mass of the fluid. Therefore, 
\begin{equation}
	\frac{dV}{dt} = -\int_0^{a(t)} J(r,t)2\pi r\ dr,
	\label{eq:balance}
\end{equation}
where the volume, $V$, is determined by the height profile: $V=\frac{1}{2} \pi H(t) a^2(t)$.

Note that Eq.~4 holds for any $J(r,t)$. In the case where the evaporation rate is constant, Eq.~4 implies
\begin{equation}
	\pi a(t) H(t) \dot a(t)+\frac{1}{2}\pi a^2(t) \dot H(t) = -\pi a^2(t) J_0,
\end{equation}
where $J(r,t)=J_0$. This differential equation is solved by employing the fact that $H(t)\propto a(t)$. Then,
\begin{equation}
	a(t) = a_0 \left(1-\frac{t}{t_f}\right)
	\label{eq:a}
\end{equation}
and
\begin{equation}
	H(t) = H_0 \left(1-\frac{t}{t_f}\right),
\end{equation}
where $a_0$ is the initial radius of the drop, $H_0$ is the initial height at the center, and $t_f = \frac{3 H_0}{2 J_0}$ is the final drying time.

% Find the flow field
\subsection{Fluid Velocity}
Together, the height profile and continuity equation uniquely determine the depth averaged velocity profile, $u(r,t)$.  After substituting $h(r,t)$ into Eq.~2, the velocity is given by
\begin{equation}
  u(r,t) =\left\{
     \begin{array}{lr}
        - \frac{r}{4(t_f-t)} &  r\leq a(t)\\
       0 &  r>a(t)
     \end{array}
   \right.
\end{equation}
The functional form of $u(r,t)$ immediately provides interesting results. In contrast to the pinned drop case, $u$ is inward and has a maximum value which is independent of time. At any time, the maximum velocity of a fluid parcel in the drop is $u(a(t),t)=-a_0/(4t_f)$. This maximal fluid velocity is slower than the rate that the edge recedes, $\dot a(t) = -a_0/t_f$. Therefore, the edge of the drop overtakes every fluid parcel that originates at a non-zero radius before the drop finishes drying. After a fluid parcel is overtaken, its mass deposits onto the substrate and remains immobile throughout the remainder of the evaporation. This deposition mechanism is qualitatively different from that for pinned drops.

%%%%%%%%%%%%%%%%%%%
% SURFACE DENSITY %
%%%%%%%%%%%%%%%%%%%

\section{Surface Density Profile}

%%%%%%%%%%%%%%%%%%%%
% DEFINITION TABLE %
%%%%%%%%%%%%%%%%%%%%
\begin{table}[tb]
	\begin{tabular}{c p{.4\textwidth}}
		\hline\\
		Symbol & Definition\\
		\hline \\
		$R$ & Final deposition radius and Lagrangian label for a mass of solute\\
		$M(R)$ & Mass bound by inner radius $R$ after deposition is complete\\
		$\tau(R)$ & Deposition time; $a(\tau(R))=R$\\ 
		$\mu(R)$ & Initial radius bounding $M(R)$ \\
		$\xi(R,t)$ & Trajectory of bounding radius over time; $\xi(R,0)=\mu(R)$ and $\xi(R,\tau(R))=R$\\
		$u(r,t)$ & Radial velocity of fluid at radius $r$ and time $t$\\
		$\eta(r)$ & Deposited surface density profile\\
		\hline
	\end{tabular}
	\caption{Definitions of symbols}
	\label{tab:def}
\end{table}
%%%%%%%%%%%%%%%%%%%%

\begin{figure}[htbp]
   \centering
   \begin{tabular}{|c|}
   \hline
   \includegraphics[width=.375\textwidth]{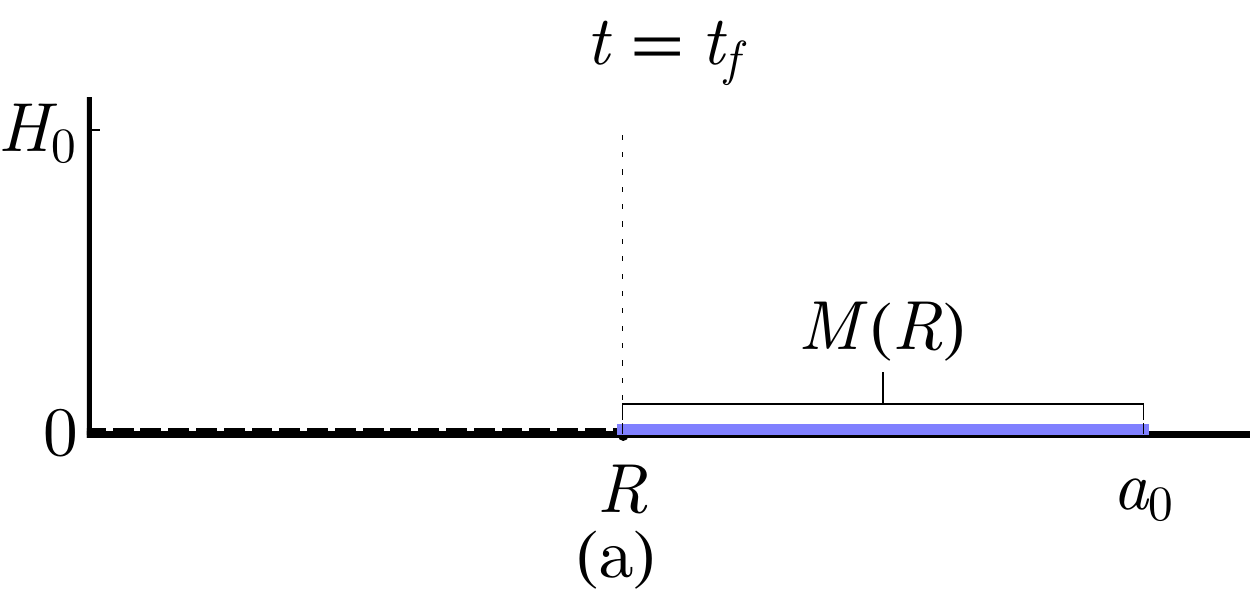}\\
   \hline
   \includegraphics[width=.375\textwidth]{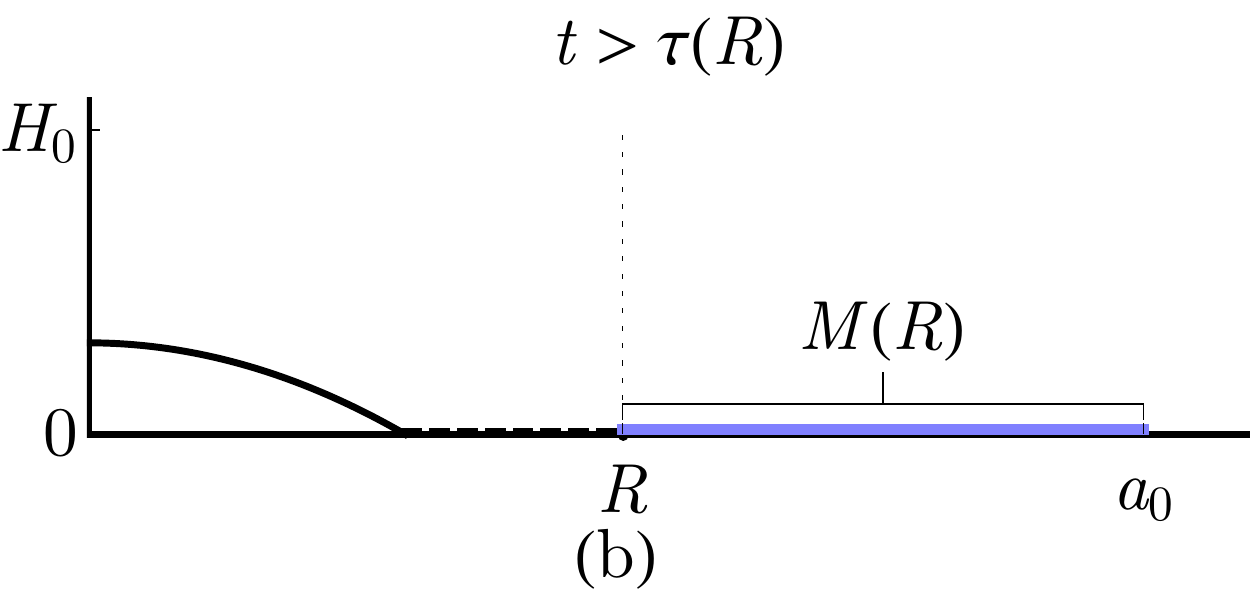}\\
   \hline
   \includegraphics[width=.375\textwidth]{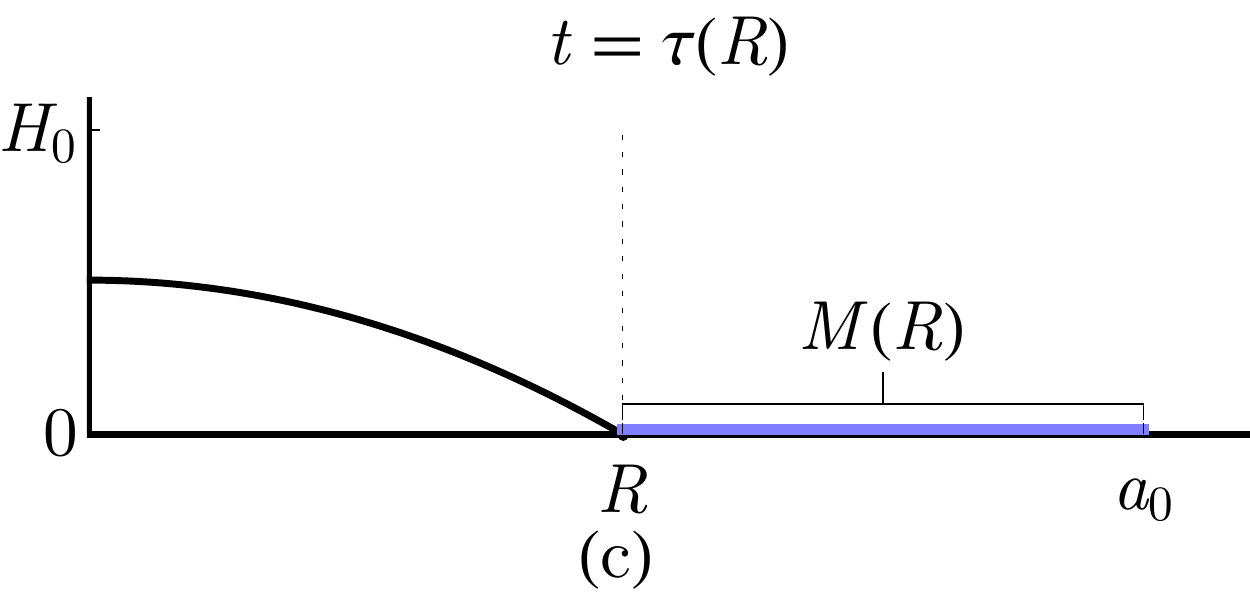}\\
   \hline
   \includegraphics[width=.375\textwidth]{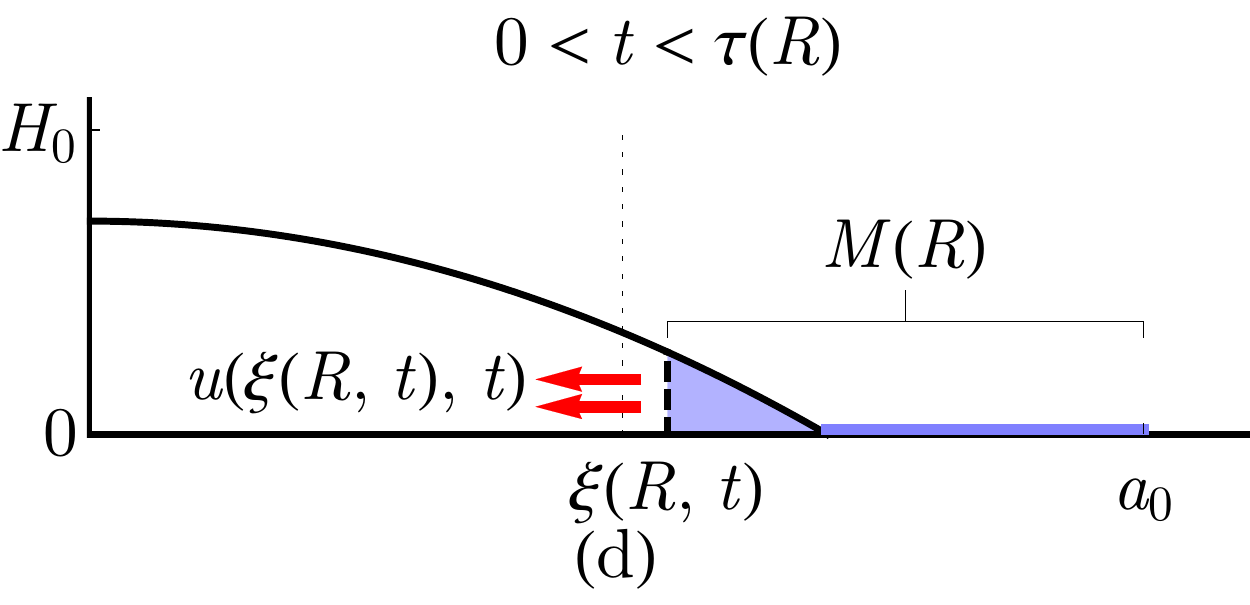}\\
   \hline
   \includegraphics[width=.375\textwidth]{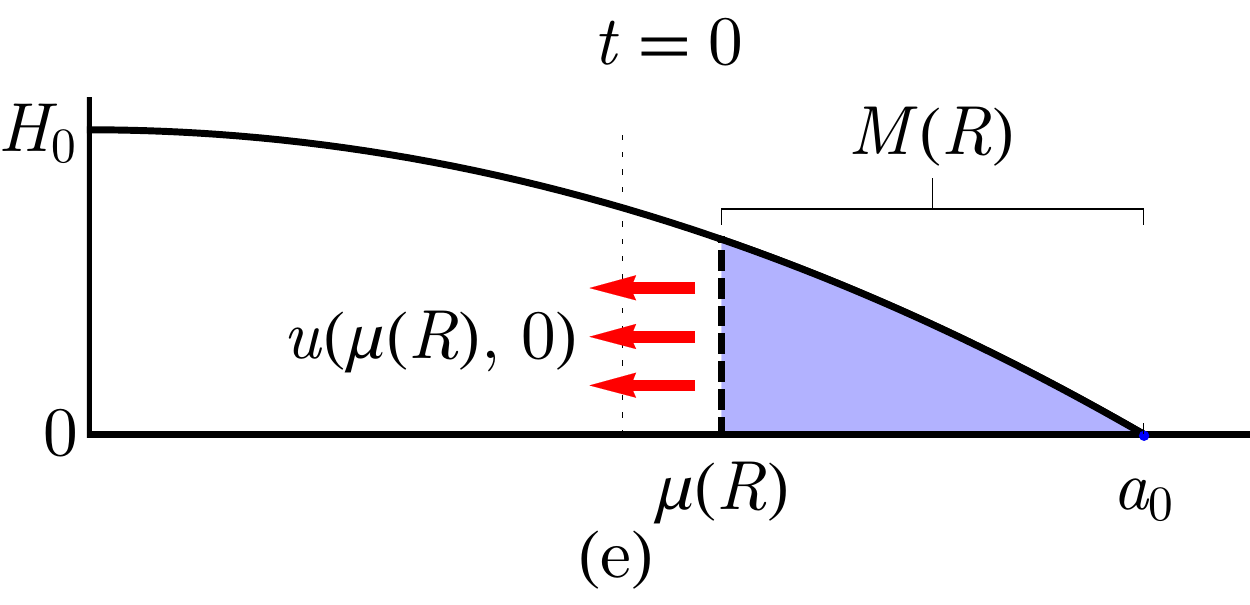}\\
   \hline
   \end{tabular}
   \caption{
   Schematic of solute deposition from evaporating drop with receding contact line. (a) through (e) shows deposition in reverse chronological order. (a) After drop has dried, all solute has deposited on the substrate. $M(R)$ is the total mass bound between radius $R$ and outer radius of the stain. (b) When the drop edge, $a(t)$, is less than $R$, the mass $M(R)$ has deposited on substrate and remains immobile throughout the remainder of the evaporation. (c) At time $\tau(R)$, $a(t)=R$ and the last parcel of $M(R)$ deposits onto substrate. (d) When $a(t) > R$, $M(R)$ continuously deposits across substrate as drop dries. Its bounding radius, $\xi(R,t)$ evolves over time. (e) Initial bounding radius, $\mu(R)=\xi(R,0)$, and height profile, $h(r,0)$, determine $M(R)$.
   }
   \label{fig:deposition}
\end{figure}

The evolution of the initial mass of solute determines the final surface density profile, $\eta(r)$. After the drop dries, the mass $M(R)$ deposited on the substrate between a radius $R$ and the initial edge of the drop $a_0$ is the integral over the surface density:
\begin{equation}
M(R) = \int_R^{a_0} 2\pi r\eta(r)dr,
\end{equation}
so that $M'(R)\equiv 2\pi R \eta(R)$. At time $t=0$, $M(R)$ can also be calculated from the initial height profile. If $\phi_0$ is the initial density of solute and $\mu(R)$ is the radius that initially bounds the mass $M(R)$,
\begin{equation}
M(R) = \int_{\mu(R)}^{a_0} 2\pi r \phi_0h(r,0)dr.
\end{equation}
(See Fig.~\ref{fig:deposition}) Then, $M'(R)$ explicitly connects $\eta$, $h$, and $\mu$:
\begin{equation}
	\eta(R) =\phi_0h(\mu(R),0)\frac{\mu(R)}{R}\mu'(R).
	\label{eq:eta}
\end{equation}
Interpreted piece by piece, this is a very intuitive equation. The annular parcel of fluid that deposits at $R$ determines the final surface density. The parcel originated at $\mu(R)$ with a local area density of $\phi_0h(\mu(R),0)$. The circumference of the parcel decreases after being transported from $\mu(R)$ to $R$, leading to an increase in density by a factor of $\mu(R)/R$. The annular parcel is also be compressed or extended radially during its evolution, which further alters its density by a factor of $\mu'(R)$. Therefore, $\mu(R)$ determines the final surface density.

%%%%%%%%%%%%%%%%%%%%%%%%
% DEPOSITION SCHEMATIC %
%%%%%%%%%%%%%%%%%%%%%%%%
%%%%%%%%%%%%%%%%%%%%%%%%

We track the evolution of an annular fluid parcel that deposits at $R$ backwards in time to find its initial radius. Let $\xi(R,t)$ be the trajectory of a fluid parcel that finally deposits at $R$. The parcel will deposit on the surface at some time $\tau(R)$. By definition, $\xi(R,\tau(R))=R$ at this time. Since a fluid parcel deposits after the receding edge overtakes it, the parcel deposits when $a(\tau(R))=R$. Substituting this condition into Eq.~6 yields
\begin{equation}
	\tau(R) =t_f(1- R/a_0).
	\label{eq:tau}
\end{equation}

\begin{figure}[tb]
   \centering
   \includegraphics[width=.4\textwidth]{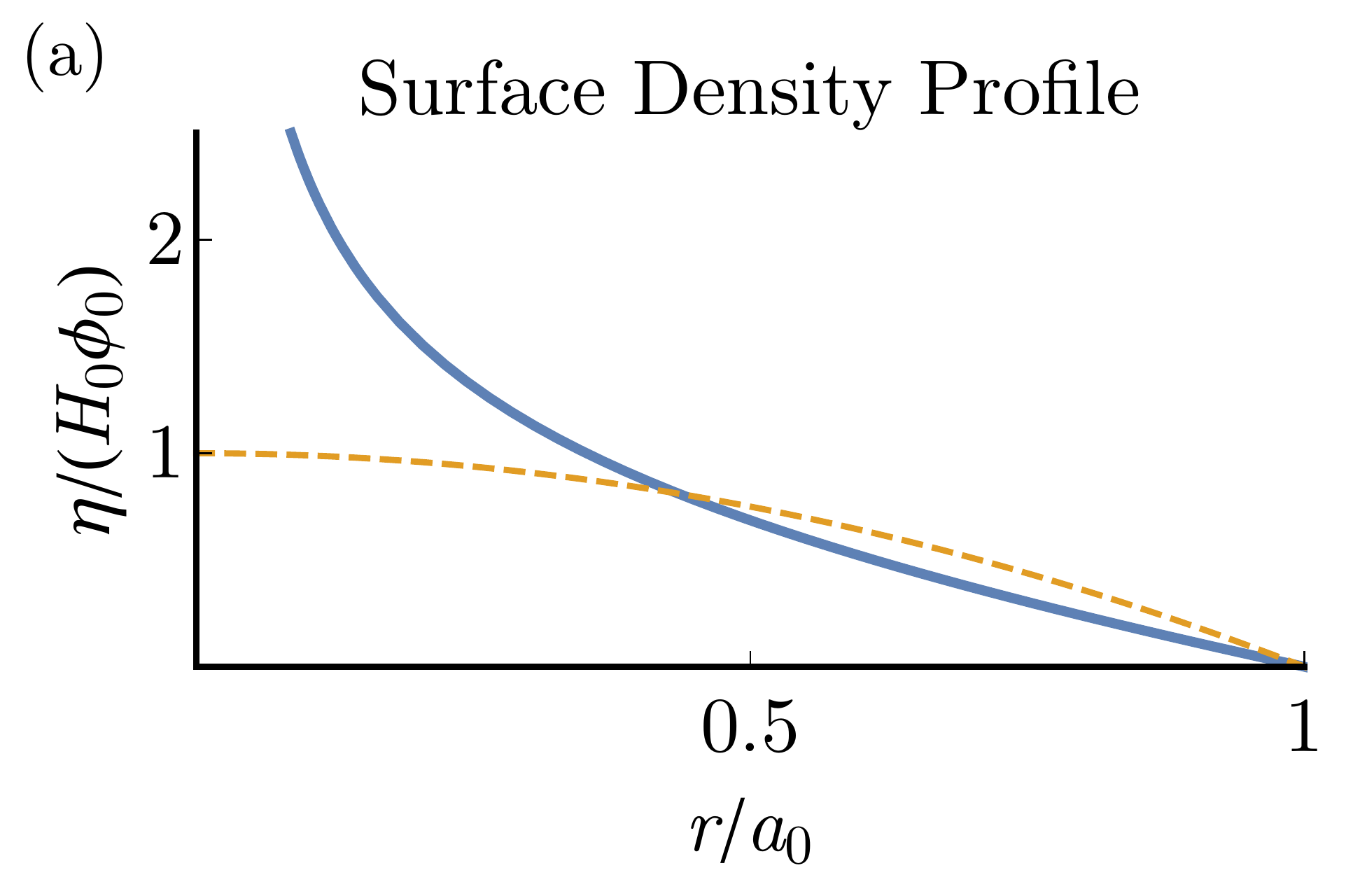}
   \includegraphics[width=.4\textwidth]{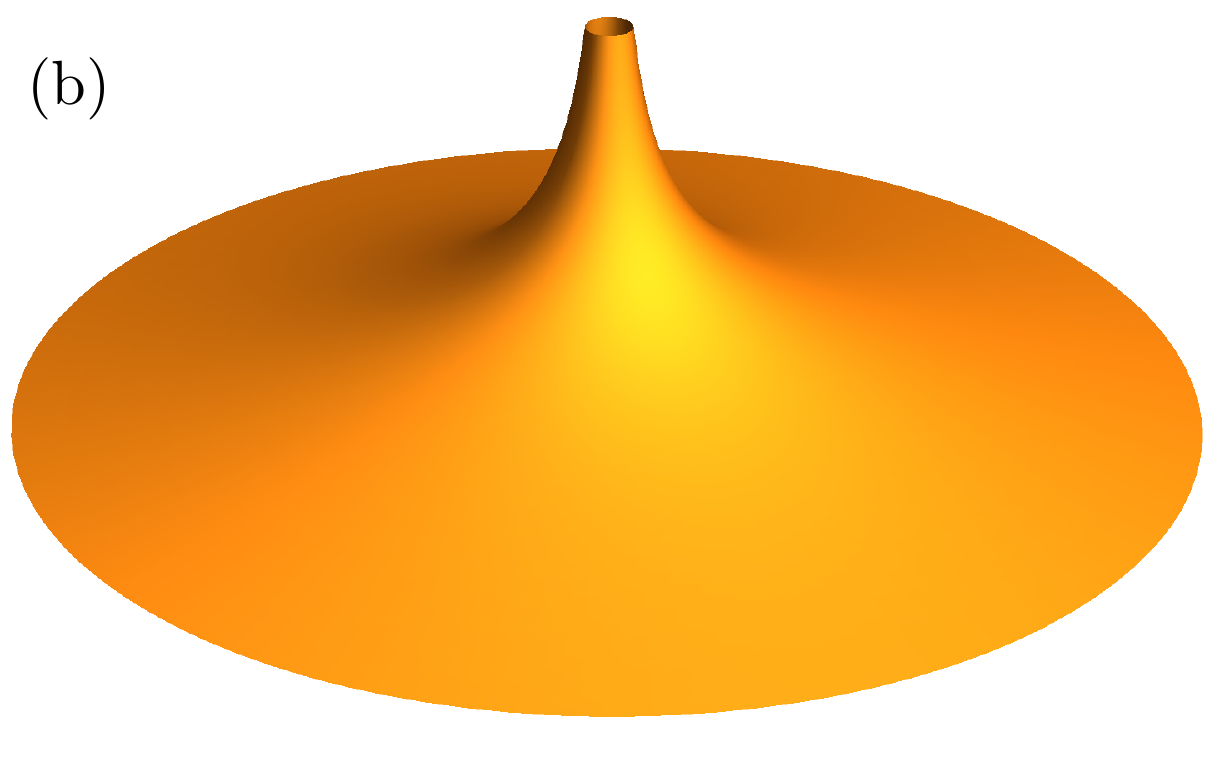}
   \caption{
   (a) Deposited surface density profile (blue, solid) compared with $\phi_0 h(r,0)$, the initial surface density (yellow, dashed). Note that the mass is concentrated and the surface density diverges at the center. The stain is shaped like a mountain instead of a ring. (b) Perspective view of the deposited stain.
   }
   \label{fig:eta}
\end{figure}

The solute moves with the local fluid velocity because it is passively transported by the fluid. Thus, the trajectory of a fluid parcel is
\begin{equation}
	\dot \xi(R,t) = u(\xi(R,t), t) = -\frac{\xi(R,t)}{4(t_f-t)}, 
	\label{eq:dxi}
\end{equation}
so that $d\xi/\xi = -dt/(4(t_f-t)),$ implying $\xi \propto (1-t/t_f)^{1/4}$. Because the parcel reaches the edge at radius $R$ and time $\tau(R)$, its trajectory is subject to the boundary condition $\xi(R,\tau(R)) = R.$ Then,
\begin{equation}
	\xi(R,t) = a_0 \left(\frac{R}{a_0}\right)^{3/4} (1-t/t_f)^{1/4}
\end{equation}
and
\begin{equation}
	\mu(R)=\xi(R,0)=a_0 \left(\frac{R}{a_0}\right)^{3/4}
	\label{eq:mu}
\end{equation}
As shown in Eq.~11, the surface density is completely determined by $\mu$. Therefore,
\begin{equation}
	\eta(r) = \frac{3}{4} \phi_0 H_0 \left(\left(\frac{r}{a_0}\right)^{-1/2}-\frac{r}{a_0}\right).\label{eq:analyticresult}
\end{equation}
See Fig.~\ref{fig:eta} for a plot of $\eta(r)$.

The nature of the divergence at small $r$ can be found quite generally. As an example, we compute the asymptotic behavior of $\eta$ for any evaporation profile $J(r,t)$ that is only a function of $r/a(t)$. In these cases $u(r,t)$ varies linearly near the center of the drop as $u(r,t) \approx \nu r/(t_f-t)$, where the dimensionless coefficient $\nu$ depends on $J$ and $h$ as discussed below. Then, Eq.~13 implies that $\xi(R,t) \approx a_0 (R/a_0)^{1-\nu} (1-t/t_f)^{-\nu}$, which further implies that $\mu\propto (R/a_0)^{1+\nu}$ and $\eta\propto (R/a_0)^{2\nu}$.

%%%%%%%%%%%%%%%
% ETA PROFILE %
%%%%%%%%%%%%%%%
%%%%%%%%%%%%%%%

%%%%%%%%%%%%%%
% DISCUSSION %
%%%%%%%%%%%%%%

\section{Discussion}

% Mountain/Divergence
The first thing to note about the surface density profile (Eq.~16) is that mass is concentrated at the center and fades towards the outside of the drop. The overall shape is more like a mountain than a ring. In fact, the surface density diverges as $(r/a_0)^{-1/2}$ towards the center. This is a weak divergence and the total mass of the stain, $M(0)$, is finite. 
% Cut off
For experiments, the divergence is cut off because one of the assumptions made in this calculation will fail at late times. For example, at late times the concentration of the solute in the fluid will increase until the solute starts interacting with itself. After this point, the assumption that the solute is a dilute passive tracer is no longer valid and the divergence will not be realized. However, the exact time when the calculation fails depends on the experimental realization. For the given example, lowering the initial solute concentration causes the assumptions to remain valid for longer times.

% Power Law
Remarkably, the power-law governing the deposit shape is controlled by experimental conditions.
As noted above, only the behavior of $u(r)$ near the center is relevant for determining the density profile $\eta(r)$. The exponent $\nu$ can
% If the velocity varies linearly as $\nu r/(t_f-t)$ near the center of the drop, the power of the divergence is $\eta(r)\propto(r/a_0)^{2\nu}$. Here, $\nu$ is the effective ``Poisson ratio'' of the stagnation point, as defined by Witten.\cite{witten2009robust} The Poisson ratio
be easily calculated from the evaporation and height profiles without explicitly solving for the complete flow field:
\begin{equation}
   \nu=\frac{1}{2}\left(1-\frac{J(0)/\bar{J}}{\dot{h}(0)/\dot{\bar{h}}}\right),
\end{equation}
where the overbar indicates the average over the drop.\cite{witten2009robust} This formula depends only on the existence of a stationary and regular stagnation point; thus it is applicable to unpinned as well as pinned circular drops. The unpinned aspect only influences the $\dot h(0)/\dot{\bar{h}}$ factor (See Fig.~\ref{fig:height}). When $\nu>0$, fluid flows away from the center and the surface density will fade to 0. For $\nu<0$, the density diverges and the stain forms a mountain. For a drop evaporating on a dry substrate, $\nu$ can be calculated analytically. The evaporation profile for a drop on a thin dry substrate is $J(r,t)=J_0 f(\lambda) \left(1-\left(r/a(t)\right)^2\right)^{-\lambda}$,
where $\lambda = (\pi-2\theta_c)/(2\pi-2\theta_c)$.\cite{deegan2000contact} In the limit where $\theta_c$ approaches 0, $\nu =1/8$. In this case, the stain fades at the center even though the contact line recedes.

% Extension
It is also worth noting that Eqs.~2, 3, 4, and 11 are generic for a thin drop with a receding contact line. Even though the functional form of the evaporation controls the velocity profile, the time dependence of the height profile, and the final form of the surface density, it is possible to follow the procedure outlined in this paper to find the surface density. In special cases, the surface density can be found analytically, but it is not difficult to extend this analysis numerically to explore the morphology induced by other evaporation profiles.

%%%%%%%%%%%%%%
% CONCLUSION %
%%%%%%%%%%%%%%

\section{Conclusion}

The fluid motion within a drying drop is directly influenced by the behavior of its contact line. Unlike a pinned drop, a receding contact line pushes fluid inwards and the stain deposits continuously as the drop evaporates. For a circular drop with Poisson ratio $\nu$ at its center, the power law of the surface density profile near the center is $\eta\propto (r/a_0)^{2\nu}$. To apply this framework accurately to an experiment requires an accurate estimate of the surface evaporation. It is not clear, for example, whether deposited mass will retain moisture, which would greatly reduce the evaporation from the edges of the drying drop. Once an evaporation profile is obtained, our method can quickly predict the profile of the deposited stain.

\section*{Acknowledgments}
The author is grateful to Efi Efrati for fruitful discussions. This work is a PhD research project supervised by T. A. Witten. It was supported in part by the University of Chicago MRSEC program of the US National Science Foundation under Award Number DMR 0820054.

\footnotesize{
\bibliography{./references.bib}

\providecommand*{\mcitethebibliography}{\thebibliography}
\csname @ifundefined\endcsname{endmcitethebibliography}
{\let\endmcitethebibliography\endthebibliography}{}
\begin{mcitethebibliography}{26}
\providecommand*{\natexlab}[1]{#1}
\providecommand*{\mciteSetBstSublistMode}[1]{}
\providecommand*{\mciteSetBstMaxWidthForm}[2]{}
\providecommand*{\mciteBstWouldAddEndPuncttrue}
  {\def\EndOfBibitem{\unskip.}}
\providecommand*{\mciteBstWouldAddEndPunctfalse}
  {\let\EndOfBibitem\relax}
\providecommand*{\mciteSetBstMidEndSepPunct}[3]{}
\providecommand*{\mciteSetBstSublistLabelBeginEnd}[3]{}
\providecommand*{\EndOfBibitem}{}
\mciteSetBstSublistMode{f}
\mciteSetBstMaxWidthForm{subitem}
{(\emph{\alph{mcitesubitemcount}})}
\mciteSetBstSublistLabelBeginEnd{\mcitemaxwidthsubitemform\space}
{\relax}{\relax}

\bibitem[Abkarian \emph{et~al.}(2004)Abkarian, Nunes, and
  Stone]{abkarian2004colloidal}
M.~Abkarian, J.~Nunes and H.~A. Stone, \emph{J. Am. Chem. Soc.}, 2004,
  \textbf{126}, 5978--5979\relax
\mciteBstWouldAddEndPuncttrue
\mciteSetBstMidEndSepPunct{\mcitedefaultmidpunct}
{\mcitedefaultendpunct}{\mcitedefaultseppunct}\relax
\EndOfBibitem
\bibitem[Nobile \emph{et~al.}(2009)Nobile, Carbone, Fiore, Cingolani, Manna,
  and Krahne]{nobile2009self}
C.~Nobile, L.~Carbone, A.~Fiore, R.~Cingolani, L.~Manna and R.~Krahne, \emph{J.
  Phys.: Condens. Matter}, 2009, \textbf{21}, 264013\relax
\mciteBstWouldAddEndPuncttrue
\mciteSetBstMidEndSepPunct{\mcitedefaultmidpunct}
{\mcitedefaultendpunct}{\mcitedefaultseppunct}\relax
\EndOfBibitem
\bibitem[Byun \emph{et~al.}(2010)Byun, Bowden, and Lin]{byun2010hierarchically}
M.~Byun, N.~B. Bowden and Z.~Lin, \emph{Nano Lett.}, 2010, \textbf{10},
  3111--3117\relax
\mciteBstWouldAddEndPuncttrue
\mciteSetBstMidEndSepPunct{\mcitedefaultmidpunct}
{\mcitedefaultendpunct}{\mcitedefaultseppunct}\relax
\EndOfBibitem
\bibitem[Mar{\'\i}n \emph{et~al.}(2011)Mar{\'\i}n, Gelderblom, Lohse, and
  Snoeijer]{marin2011order}
{\'A}.~G. Mar{\'\i}n, H.~Gelderblom, D.~Lohse and J.~H. Snoeijer, \emph{Phys.
  Rev. Lett.}, 2011, \textbf{107}, 085502\relax
\mciteBstWouldAddEndPuncttrue
\mciteSetBstMidEndSepPunct{\mcitedefaultmidpunct}
{\mcitedefaultendpunct}{\mcitedefaultseppunct}\relax
\EndOfBibitem
\bibitem[Li \emph{et~al.}(2013)Li, Han, Byun, Zhu, Zou, and
  Lin]{li2013macroscopic}
B.~Li, W.~Han, M.~Byun, L.~Zhu, Q.~Zou and Z.~Lin, \emph{ACS Nano}, 2013,
  \textbf{7}, 4326--4333\relax
\mciteBstWouldAddEndPuncttrue
\mciteSetBstMidEndSepPunct{\mcitedefaultmidpunct}
{\mcitedefaultendpunct}{\mcitedefaultseppunct}\relax
\EndOfBibitem
\bibitem[Kim \emph{et~al.}(2006)Kim, Jeong, Park, and Moon]{kim2006direct}
D.~Kim, S.~Jeong, B.~K. Park and J.~Moon, \emph{Appl. Phys. Lett.}, 2006,
  \textbf{89}, 264101\relax
\mciteBstWouldAddEndPuncttrue
\mciteSetBstMidEndSepPunct{\mcitedefaultmidpunct}
{\mcitedefaultendpunct}{\mcitedefaultseppunct}\relax
\EndOfBibitem
\bibitem[Park \emph{et~al.}(2014)Park, Lee, Kim, Kim, Lee, Park, Lee, and
  Cho]{park2014flexible}
J.~H. Park, D.~Y. Lee, Y.-H. Kim, J.~K. Kim, J.~H. Lee, J.~H. Park, T.-W. Lee
  and J.~H. Cho, \emph{ACS Appl. Mater. Interfaces}, 2014, \textbf{6},
  12380--12387\relax
\mciteBstWouldAddEndPuncttrue
\mciteSetBstMidEndSepPunct{\mcitedefaultmidpunct}
{\mcitedefaultendpunct}{\mcitedefaultseppunct}\relax
\EndOfBibitem
\bibitem[Monteux and Lequeux(2011)]{monteux2011packing}
C.~Monteux and F.~Lequeux, \emph{Langmuir}, 2011, \textbf{27}, 2917--2922\relax
\mciteBstWouldAddEndPuncttrue
\mciteSetBstMidEndSepPunct{\mcitedefaultmidpunct}
{\mcitedefaultendpunct}{\mcitedefaultseppunct}\relax
\EndOfBibitem
\bibitem[Brutin \emph{et~al.}(2011)Brutin, Sobac, Loquet, and
  Sampol]{brutin2011pattern}
D.~Brutin, B.~Sobac, B.~Loquet and J.~Sampol, \emph{J. Fluid Mech.}, 2011,
  \textbf{667}, 85--95\relax
\mciteBstWouldAddEndPuncttrue
\mciteSetBstMidEndSepPunct{\mcitedefaultmidpunct}
{\mcitedefaultendpunct}{\mcitedefaultseppunct}\relax
\EndOfBibitem
\bibitem[Hu and Larson(2006)]{hu2006marangoni}
H.~Hu and R.~G. Larson, \emph{J. Phys. Chem. B}, 2006, \textbf{110},
  7090--7094\relax
\mciteBstWouldAddEndPuncttrue
\mciteSetBstMidEndSepPunct{\mcitedefaultmidpunct}
{\mcitedefaultendpunct}{\mcitedefaultseppunct}\relax
\EndOfBibitem
\bibitem[Xu \emph{et~al.}(2007)Xu, Xia, and Lin]{xu2007evaporation}
J.~Xu, J.~Xia and Z.~Lin, \emph{Angew. Chem. Int. Ed.}, 2007, \textbf{119},
  1892--1895\relax
\mciteBstWouldAddEndPuncttrue
\mciteSetBstMidEndSepPunct{\mcitedefaultmidpunct}
{\mcitedefaultendpunct}{\mcitedefaultseppunct}\relax
\EndOfBibitem
\bibitem[Hong \emph{et~al.}(2009)Hong, Byun, and Lin]{hong2009robust}
S.~W. Hong, M.~Byun and Z.~Lin, \emph{Angew. Chem. Int. Ed.}, 2009,
  \textbf{48}, 512--516\relax
\mciteBstWouldAddEndPuncttrue
\mciteSetBstMidEndSepPunct{\mcitedefaultmidpunct}
{\mcitedefaultendpunct}{\mcitedefaultseppunct}\relax
\EndOfBibitem
\bibitem[Yunker \emph{et~al.}(2011)Yunker, Still, Lohr, and
  Yodh]{yunker2011suppression}
P.~J. Yunker, T.~Still, M.~A. Lohr and A.~Yodh, \emph{Nature}, 2011,
  \textbf{476}, 308--311\relax
\mciteBstWouldAddEndPuncttrue
\mciteSetBstMidEndSepPunct{\mcitedefaultmidpunct}
{\mcitedefaultendpunct}{\mcitedefaultseppunct}\relax
\EndOfBibitem
\bibitem[Deegan \emph{et~al.}(1997)Deegan, Bakajin, Dupont, Huber, Nagel, and
  Witten]{deegan1997capillary}
R.~D. Deegan, O.~Bakajin, T.~F. Dupont, G.~Huber, S.~R. Nagel and T.~A. Witten,
  \emph{Nature}, 1997, \textbf{389}, 827--829\relax
\mciteBstWouldAddEndPuncttrue
\mciteSetBstMidEndSepPunct{\mcitedefaultmidpunct}
{\mcitedefaultendpunct}{\mcitedefaultseppunct}\relax
\EndOfBibitem
\bibitem[Deegan \emph{et~al.}(2000)Deegan, Bakajin, Dupont, Huber, Nagel, and
  Witten]{deegan2000contact}
R.~D. Deegan, O.~Bakajin, T.~F. Dupont, G.~Huber, S.~R. Nagel and T.~A. Witten,
  \emph{Phys. Rev. E}, 2000, \textbf{62}, 756--765\relax
\mciteBstWouldAddEndPuncttrue
\mciteSetBstMidEndSepPunct{\mcitedefaultmidpunct}
{\mcitedefaultendpunct}{\mcitedefaultseppunct}\relax
\EndOfBibitem
\bibitem[Deegan(2000)]{deegan2000pattern}
R.~D. Deegan, \emph{Phys. Rev. E}, 2000, \textbf{61}, 475--485\relax
\mciteBstWouldAddEndPuncttrue
\mciteSetBstMidEndSepPunct{\mcitedefaultmidpunct}
{\mcitedefaultendpunct}{\mcitedefaultseppunct}\relax
\EndOfBibitem
\bibitem[Yabu and Shimomura(2005)]{yabu2005preparation}
H.~Yabu and M.~Shimomura, \emph{Adv. Funct. Mater.}, 2005, \textbf{15},
  575--581\relax
\mciteBstWouldAddEndPuncttrue
\mciteSetBstMidEndSepPunct{\mcitedefaultmidpunct}
{\mcitedefaultendpunct}{\mcitedefaultseppunct}\relax
\EndOfBibitem
\bibitem[Maheshwari \emph{et~al.}(2008)Maheshwari, Zhang, Zhu, and
  Chang]{maheshwari2008coupling}
S.~Maheshwari, L.~Zhang, Y.~Zhu and H.-C. Chang, \emph{Phys. Rev. Lett.}, 2008,
  \textbf{100}, 044503\relax
\mciteBstWouldAddEndPuncttrue
\mciteSetBstMidEndSepPunct{\mcitedefaultmidpunct}
{\mcitedefaultendpunct}{\mcitedefaultseppunct}\relax
\EndOfBibitem
\bibitem[Frastia \emph{et~al.}(2011)Frastia, Archer, and
  Thiele]{frastia2011dynamical}
L.~Frastia, A.~J. Archer and U.~Thiele, \emph{Phys. Rev. Lett.}, 2011,
  \textbf{106}, 077801\relax
\mciteBstWouldAddEndPuncttrue
\mciteSetBstMidEndSepPunct{\mcitedefaultmidpunct}
{\mcitedefaultendpunct}{\mcitedefaultseppunct}\relax
\EndOfBibitem
\bibitem[Mampallil \emph{et~al.}(2012)Mampallil, Eral, van~den Ende, and
  Mugele]{mampallil2012control}
D.~Mampallil, H.~Eral, D.~van~den Ende and F.~Mugele, \emph{Soft Matter}, 2012,
  \textbf{8}, 10614--10617\relax
\mciteBstWouldAddEndPuncttrue
\mciteSetBstMidEndSepPunct{\mcitedefaultmidpunct}
{\mcitedefaultendpunct}{\mcitedefaultseppunct}\relax
\EndOfBibitem
\bibitem[Zhang \emph{et~al.}(2014)Zhang, Nguyen, and Chen]{zhang2014coffee}
L.~Zhang, Y.~Nguyen and W.~Chen, \emph{Colloids Surf., A}, 2014, \textbf{449},
  42--50\relax
\mciteBstWouldAddEndPuncttrue
\mciteSetBstMidEndSepPunct{\mcitedefaultmidpunct}
{\mcitedefaultendpunct}{\mcitedefaultseppunct}\relax
\EndOfBibitem
\bibitem[Li \emph{et~al.}(2014)Li, Sheng, and Tsao]{li2014solute}
Y.-F. Li, Y.-J. Sheng and H.-K. Tsao, \emph{Langmuir}, 2014, \textbf{30},
  7716--7723\relax
\mciteBstWouldAddEndPuncttrue
\mciteSetBstMidEndSepPunct{\mcitedefaultmidpunct}
{\mcitedefaultendpunct}{\mcitedefaultseppunct}\relax
\EndOfBibitem
\bibitem[Willmer \emph{et~al.}(2010)Willmer, Baldwin, Kwartnik, and
  Fairhurst]{willmer2010growth}
D.~Willmer, K.~A. Baldwin, C.~Kwartnik and D.~J. Fairhurst, \emph{Phys. Chem.
  Chem. Phys.}, 2010, \textbf{12}, 3998--4004\relax
\mciteBstWouldAddEndPuncttrue
\mciteSetBstMidEndSepPunct{\mcitedefaultmidpunct}
{\mcitedefaultendpunct}{\mcitedefaultseppunct}\relax
\EndOfBibitem
\bibitem[Baldwin \emph{et~al.}(2012)Baldwin, Roest, Fairhurst, Sefiane, and
  Shanahan]{baldwin2012monolith}
K.~A. Baldwin, S.~Roest, D.~J. Fairhurst, K.~Sefiane and M.~E. Shanahan,
  \emph{J. Fluid Mech.}, 2012, \textbf{695}, 321--329\relax
\mciteBstWouldAddEndPuncttrue
\mciteSetBstMidEndSepPunct{\mcitedefaultmidpunct}
{\mcitedefaultendpunct}{\mcitedefaultseppunct}\relax
\EndOfBibitem
\bibitem[Drelich \emph{et~al.}(1996)Drelich, Miller, and
  Good]{drelich1996effect}
J.~Drelich, J.~D. Miller and R.~J. Good, \emph{J. Colloid Interface Sci.},
  1996, \textbf{179}, 37--50\relax
\mciteBstWouldAddEndPuncttrue
\mciteSetBstMidEndSepPunct{\mcitedefaultmidpunct}
{\mcitedefaultendpunct}{\mcitedefaultseppunct}\relax
\EndOfBibitem
\bibitem[Witten(2009)]{witten2009robust}
T.~Witten, \emph{Europhys. Lett.}, 2009, \textbf{86}, 64002\relax
\mciteBstWouldAddEndPuncttrue
\mciteSetBstMidEndSepPunct{\mcitedefaultmidpunct}
{\mcitedefaultendpunct}{\mcitedefaultseppunct}\relax
\EndOfBibitem
\end{mcitethebibliography}
\bibliographystyle{rsc} %the RSC's .bst file
}

\end{document}